\documentclass[11pt]{article}
\usepackage[utf8]{inputenc}
\usepackage[T1]{fontenc}
\usepackage[a4paper]{geometry}
\usepackage{amsmath,amsfonts,amssymb}
\usepackage{mathrsfs}
\usepackage{theorem}
\usepackage{color}
\usepackage[square,comma,sort&compress]{natbib}
\usepackage{hyperref}

\newtheorem{thm}{Theorem}[section]
\newtheorem{prop}[thm]{Proposition}

\theorembodyfont{\rmfamily}
\newtheorem{rmk}{Remark}[section]

\newcommand{\CC}{{\mathbb C}}

\newcommand{\NN}{{\mathbb N}}

\newcommand{\wt}[1]{\widetilde{#1}}

\newcommand{\half}{{\textstyle{\frac{1}{2}}}}
\newcommand{\sfrac}[2]{{\textstyle{\frac{#1}{#2}}}}
\newcommand{\proof}{\textbf{Proof:}~}

\renewcommand{\author}[1]{\normalsize\bf\raggedright#1 \\}

\DeclareMathOperator{\id}{id}

\numberwithin{equation}{section}

\begin{document}
\pagestyle{empty}

\null
\vspace{20pt}

\parskip=6pt

\begin{center}
\begin{LARGE}
\textbf{The $q$-Higgs and Askey--Wilson algebras}
\end{LARGE}

\vspace{50pt}

\begin{large}
{L.~Frappat${}^a$\footnote{luc.frappat@lapth.cnrs.fr}, J.~Gaboriaud${}^b$\footnote{gaboriaud@CRM.UMontreal.CA}, E.~Ragoucy${}^a$\footnote{eric.ragoucy@lapth.cnrs.fr}, L.~Vinet${}^b$\footnote{vinet@CRM.UMontreal.CA} 
}
\end{large}

\vspace{15mm}

${}^a$ \textit{Laboratoire d'Annecy-le-Vieux de Physique Th\'eorique LAPTh, \\ BP 110 Annecy-le-Vieux, F-74941 Annecy Cedex, France. \\
Univ. Grenoble Alpes, Univ. Savoie Mont Blanc, CNRS, F-74000 Annecy, France.}

\vspace{5mm}

${}^b$ \textit{Centre de Recherches Math\'ematiques, Universit\'e de Montr\'eal, \\
P.O. Box 6128, Centre-ville Station, Montr\'eal (Qu\'ebec), H3C 3J7, Canada.}

\end{center}

\vspace{4mm}

\begin{abstract}
A $q$-analogue of the Higgs algebra, which describes the symmetry properties of the harmonic oscillator on the $2$-sphere, is obtained as the commutant of the $\mathfrak{o}_{q^{1/2}}(2) \oplus \mathfrak{o}_{q^{1/2}}(2)$ subalgebra of $\mathfrak{o}_{q^{1/2}}(4)$ in the $q$-oscillator representation of the quantized universal enveloping algebra $U_q(\mathfrak{u}(4))$. This $q$-Higgs algebra is also found as a specialization of the Askey--Wilson algebra embedded in the tensor product $U_q(\mathfrak{su}(1,1))\otimes U_q(\mathfrak{su}(1,1))$. The connection between these two approaches is established on the basis of the Howe duality of the pair $\big(\mathfrak{o}_{q^{1/2}}(4),U_q(\mathfrak{su}(1,1))\big)$. 
\end{abstract}



\parindent=0pt
\pagestyle{plain}

\section*{Introduction}
The Higgs algebra was first obtained by Higgs \cite{Higgs1979} as the algebra of the conserved quantities of the Coulomb problem and harmonic oscillator on the $2$-sphere. 
Shown to be isomorphic to the Hahn algebra \cite{Vinet2018}, it was also identified as the symmetry algebra of the Hartmann potential \cite{Granovskii1991}, of certain ring-shaped potentials \cite{Granovskii1992} and of the singular oscillator in two dimensions \cite{Letourneau1995,Genest2014}.
The Higgs algebra stands between Lie algebras and quantized universal enveloping algebras, as it can be viewed both as a deformation of the $\mathfrak{su}(2)$ Lie algebra \cite{Bonatsos1995} and a truncation of the $U_q(\mathfrak{sl}_2)$ quantum algebra \cite{Zhedanov1992}.
It has been obtained as the quantum finite W-algebra $W(\mathfrak{sp}(4),2\,\mathfrak{sl}(2))$ \cite{Bowcock1994,Barbarin1995} and has also appeared in the context of Heisenberg quantization of identical particles \cite{Leinaas1993}. 

The Higgs algebra can be presented in the following form
\begin{align}\label{eq:higgsalg}
\begin{aligned}{}
 [D,A_\pm]&=\pm 4A_\pm,\\
 [A_+,A_-]&=-D^{3}+\alpha_1 D+\alpha_2,
\end{aligned}
\end{align}
where $\alpha_1$, $\alpha_2$ are central elements. 

We here aim to construct a $q$-deformation of \eqref{eq:higgsalg} that preserves the general algebraic underpinnings of this structure. This will lead to an algebra that differs from the one  in \cite{Chung2014} where a certain $q$-extension of the Higgs algebra was defined by simply replacing the cubic expression in $D$ by one involving $q$-numbers (see \eqref{eq:qnum}).


We propose to obtain a $q$-analogue of the Higgs algebra by following a commmutant approach similar to \cite{Frappat2018} (see also \cite{Gaboriaud2018a,Gaboriaud2018b}), where the ordinary Higgs algebra was obtained as the commutant of the $\mathfrak{o}(2)\oplus\mathfrak{o}(2)$ subalgebra of $\mathfrak{o}(4)$ in the oscillator representation of $U(\mathfrak{u}(4))$. This characterization was shown to be in duality in the sense of Howe \cite{Howe1987,Howe1989,Howe1989a,Rowe2012} with the well-established embedding of the Hahn algebra in $U(\mathfrak{su}(1,1))\otimes U(\mathfrak{su}(1,1))$ \cite{Granovskii1988,Zhedanov1993}. While Howe duality, sometimes called ``complementarity'', has not been thoroughly studied in the context of $q$-algebras (see for instance \cite{Quesne1992,Smirnov1992,Green1999,Lehrer2011,Futorny2017}), the results in \cite{Noumi1996} will provide appropriate background for our purposes.
The merit of the approach we propose is that the $q$-Higgs algebra obtained as a commutant also appears in a dual fashion as a specialization of the Askey--Wilson algebra \cite{Zhedanov1991,Terwilliger2011,Huang2017} in the tensor product $U_q(\mathfrak{su}(1,1))^{\otimes 2}$. 

Let us now briefly present the contents of the paper. In Section \ref{sec:quantumalgebras}, the $q$-deformations of $\mathfrak{su}(1,1)$ and $\mathfrak{o}(n)$ (respectively denoted $U_q(\mathfrak{su}(1,1))$ and $\mathfrak{o}_q(n)$) will be introduced along with their $q$-oscillator realizations. In Section \ref{sec:qhiggs}, a $q$-deformation of the Higgs algebra will be obtained as a commutant of $\mathfrak{o}_{q^{1/2}}(2) \oplus \mathfrak{o}_{q^{1/2}}(2)\subset\mathfrak{o}_{q^{1/2}}(4)$ in the $q$-oscillator realization of $U_q(\mathfrak{u}(4))$. The embedding of a special case of the Askey--Wilson algebra into $U_q(\mathfrak{su}(1,1))^{\otimes 2}$ will be presented in Section \ref{sec:aw}. As will be shown in Section \ref{sec:dualpair}, the $q$-Higgs algebra proves to be isomorphic to that specialization of the Askey--Wilson algebra, and this result will be explained by invoking the fact that the pair $\big(\mathfrak{o}_{q^{1/2}}(4),U_q(\mathfrak{su}(1,1))\big)$ behaves as a Howe dual pair in this context. Concluding remarks and perspectives will form the last section.

\section{The $U_q(\mathfrak{su}(1,1))$, $\mathfrak{o}_q(n)$ algebras and their $q$-oscillators realizations }\label{sec:quantumalgebras}
The duality connection that we shall invoke in our discussion involves the algebras $U_q(\mathfrak{su}(1,1))$ and $\mathfrak{o}_q(n)$. We shall thus begin by introducing these algebras and their $q$-oscillator realizations.

Let $q$ be a complex number such that $|q| < 1$. One defines for any number $x$ the following $q$-numbers:
\begin{equation}\label{eq:qnum}
(x)_q := \frac{1-q^x}{1-q} \quad \text{and} \quad [x]_q := \frac{q^x-q^{-x}}{q-q^{-1}} .
\end{equation}
The same notation will be used for operators.

\subsection{The $U_q(\mathfrak{su}(1,1))$ and $\mathfrak{o}_q(n)$ quantum algebras}
$U_{q}(\mathfrak{sl}_2)$ \cite{Drinfeld1985,Jimbo1986} is the quantized universal enveloping algebra with three generators $j_0$ and $j_\pm$ subjected to the relations
\begin{equation}
\big[ j_0 \,, j_\pm \big] = \pm j_\pm , \qquad \big[ j_+ \,, j_- \big] = [2j_0]_{q} .
\end{equation}
It is endowed with a Hopf structure with coproduct $\Delta : U_{q}(\mathfrak{sl}_2) \to U_{q}(\mathfrak{sl}_2) \otimes U_{q}(\mathfrak{sl}_2)$
\begin{equation}
\Delta (j_0) = j_0 \otimes 1 + 1 \otimes j_0 , \qquad \Delta (j_+) = j_+ \otimes q^{2j_0} + 1 \otimes j_+ , \qquad \Delta (j_-) = j_- \otimes 1 + q^{-2j_0} \otimes j_- .
\end{equation}We shall denote by $U_q(\mathfrak{su}(1,1))$ the non-compact real form of $U_q(\mathfrak{sl}_2)$ that has the three generators $J_\pm$ and $J_0$ obeying 
\begin{equation}
\big[ J_0 \,, J_\pm \big] = \pm J_\pm , \qquad J_-J_+ - q^2 J_+J_- = q^{2J_0} \, \big[ 2J_0 \big]_q . 
\label{eq:relcomsu}
\end{equation}
The coproduct $\Delta : U_q(\mathfrak{su}(1,1)) \to U_q(\mathfrak{su}(1,1)) \otimes U_q(\mathfrak{su}(1,1))$ will read
\begin{equation}
\Delta (J_0) = J_0 \otimes 1 + 1 \otimes J_0 , \qquad \Delta (J_\pm) = J_\pm \otimes q^{2J_0} + 1 \otimes J_\pm .
\label{eq:coprodsu}
\end{equation}
The Casimir operator $C$ of this algebra has the following expression
\begin{equation}
C = J_+J_- q^{-2J_0+1} - \frac{q}{(q^2-1)^2} \, \big( q^{2J_0-1} + q^{-2J_0+1} \big) + \frac{q^2+1}{(q^2-1)^2} .
\label{eq:casimirsu}
\end{equation}
The coproduct being an algebra morphism, the relations \eqref{eq:coprodsu} define an embedding of $U_q(\mathfrak{su}(1,1))$ into $U_q(\mathfrak{su}(1,1)) \otimes U_q(\mathfrak{su}(1,1))$. 
\begin{rmk}
In the limit $q \to 1$, one recovers the usual $\mathfrak{su}(1,1)$ Lie algebra with Casimir operator \mbox{$C = J_+J_--{J_0}^2   + J_0$}. Moreover, the standard presentation of $U_q(\mathfrak{su}(1,1))$ \cite{Klimyk1997} is recovered if one considers instead the generators $\wt J_0=J_0$,~ $\wt J_+ = J_+q^{-J_0}$ and $\wt J_- = q^{-J_0}J_-$, which satisfy the commutation relations $\big[ \wt J_0 \,, \wt J_\pm \big] = \pm \wt J_\pm$ and $\big[ \wt J_- \,, \wt J_+ \big] = \big[ 2\wt J_0 \big]_q$ and have co-commutative coproduct.
\end{rmk}

We introduce next the non-standard $q$-deformation $\mathfrak{o}_q(n)$ of $\mathfrak{o}(n)$ which is defined as the associative unital algebra with generators $L_{i,i+1}$ ($i=1,\dots,n-1$) and relations
\begin{subequations}\label{eq:soqm}
\begin{align}
& L_{i-1,i}\,L_{i,i+1}^2 - (q+q^{-1}) L_{i,i+1}\,L_{i-1,i}\,L_{i,i+1} + L_{i,i+1}^2\,L_{i-1,i} = -L_{i-1,i} \label{eq:soqm1} , \\
& L_{i,i+1}\,L_{i-1,i}^2 - (q+q^{-1}) L_{i-1,i}\,L_{i,i+1}\,L_{i-1,i} + L_{i-1,i}^2\,L_{i,i+1} = -L_{i,i+1} \label{eq:soqm2} , \\
& \big[ L_{i,i+1} \,, L_{j,j+1} \big] = 0 \quad \text{for} \quad |i-j| > 1 . \label{eq:soqm3}
\end{align}
\end{subequations}
In the litterature, this non-standard deformation is often denoted $U'_q(\mathfrak{so}_n)$, see for instance \cite{Gavrilik2000,Klimyk2001,Klimyk2002,Iorgov2005}. 
It has been shown in \cite{Noumi1996a} that $\mathfrak{o}_q(n)$ can be viewed as a $q$-analogue of the symmetric space based on the pair $(\mathfrak{gl}(n),\mathfrak{o}(n))$.
Although it has no Hopf structure on its own, it is a coideal subalgebra of $U_q(\mathfrak{sl}(n))$ \cite{Noumi1996a} and
 appears in many areas of mathematical physics \cite{Klimyk2002}.

The two cases where $n=3$ and $n=4$ are especially of interest to us.

Let us first note that it is possible to consider a so-called ``Cartesian'' presentation \cite{Zhedanov1992a,Havlicek1999,Havlicek2001} of $U_q(\mathfrak{sl}_2)$, in which the three generators play an ``equitable'' role, and which corresponds to the non-standard deformation $\mathfrak{o}_q(3)$ (equivalently $U'_q(\mathfrak{so}_3)$ in refs. \cite{Havlicek1999,Havlicek2001}) of the universal enveloping algebra $U(\mathfrak{so}(3))$, obtained by modifying the defining relations for the skew-symmetric generators of $\mathfrak{so}(3)$.

It goes like this. With $j_0$, $j_\pm$, the $U_q(\mathfrak{sl}_2)$ generators, form the following elements:
\begin{align}
\begin{aligned}\label{eq:jiequit}
j_1 &= ig\big\{ q^{\frac{1}{2}j_0} \,,\, j_+ + j_- \big\} ,  \\
j_2 &= g\big\{ q^{-\frac{1}{2}j_0} \,,\, j_+ - j_- \big\}, 
\end{aligned}\qquad\quad g = \frac{1}{(q^{\frac{1}{4}}+q^{-\frac{1}{4}})(q^{\frac{1}{2}}+q^{-\frac{1}{2}})},
\end{align}
where $\{a,b\}=ab+ba$~ is the anticommutator and $g$ is a normalization factor. Defining $j_3 \equiv [j_1,j_2]_q$, where $[a,b]_q:= q^{\frac{1}{2}} ab-q^{-\frac{1}{2}}ba$~ is the $q$-commutator, $j_1$, $j_2$ and $j_3$ then satisfy the ``Cartesian'' relations
\begin{align}
[j_1,j_2]_q=j_3,\qquad [j_2,j_3]_q=j_1,\qquad [j_3,j_1]_q=j_2.\label{eq:relcom1}
\end{align}
Upon identifying $L_{12}=j_1$, $L_{23}=j_2$, one finds that this corresponds precisely to the relations \eqref{eq:soqm} for the algebra $\mathfrak{o}_q(3)$. Note that the relations \eqref{eq:soqm3} do not exist in this case.

For what follows, it will also be useful to have the formulas for $\mathfrak{o}_q(4)$ in full. These relations read \cite{Klimyk1994}
\begin{subequations}\label{eq:soq4}
\begin{align}
L_{12}\,L_{23}^2 - (q+q^{-1}) L_{23}\,L_{12}\,L_{23} + L_{23}^2\,L_{12} &= -L_{12} \label{eq:soq41} , \\
L_{23}\,L_{12}^2 - (q+q^{-1}) L_{12}\,L_{23}\,L_{12} + L_{12}^2\,L_{23} &= -L_{23} \label{eq:soq42} , \\
L_{23}\,L_{34}^2 - (q+q^{-1}) L_{34}\,L_{23}\,L_{34} + L_{34}^2\,L_{23} &= -L_{23} \label{eq:soq43} , \\
L_{34}\,L_{23}^2 - (q+q^{-1}) L_{23}\,L_{34}\,L_{23} + L_{23}^2\,L_{34} &= -L_{34} \label{eq:soq44} , \\
&\hspace{-9em} \big[ L_{12} \,, L_{34} \big] = 0 . \label{eq:soq45}
\end{align}
\end{subequations}
It is immediate to see that $L_{12},L_{23}$ and $L_{23},L_{34}$ respectively generate two $\mathfrak{o}_q(3)$ subalgebras of $\mathfrak{o}_q(4)$,
however they do not appear within a direct sum, in contrast to what happens with $\mathfrak{o}(4)$.

If one introduces the following elements:
\begin{equation}
L_{13}^\pm = \big[ L_{12} \,, L_{23} \big]_{q^{\pm 1}} , \qquad 
L_{24}^\pm = \big[ L_{23} \,, L_{34} \big]_{q^{\pm 1}} , \qquad
L_{14}^\pm = \big[ L_{13}^\pm \,, L_{34} \big]_{q^{\pm 1}},
\end{equation}
where $[a,b]_{q}$ is defined as above and $[a,b]_{q^{-1}} := q^{-\frac{1}{2}} ab-q^{\frac{1}{2}}ba$, the two independent Casimir operators of the algebra $\mathfrak{o}_q(4)$ are then given by \cite{Noumi1996,Gavrilik2000,Havlicek2001}
\begin{subequations}\label{eq:casimirsoq4}
\begin{align}
C_4 &= q^{-2} {L_{12}}^2 + {L_{23}}^2 + q^2 {L_{34}}^2 + q^{-1} L_{13}^+\,L_{13}^- + q L_{24}^+\,L_{24}^- + L_{14}^+\,L_{14}^- , 
\label{eq:casimir4} \\
C'_4 &= q^{-1} L_{12}\,L_{34} - L_{13}^{+}\,L_{24}^{+} + q L_{23}\,L_{14}^{+} .
\label{eq:casimir4p}
\end{align}
\end{subequations}

\subsection{The $q$-oscillator algebras, Schwinger and metaplectic realizations}
Let us now recall the properties of the $q$-oscillator operators that will be used to realize the algebras presented above. The $q$-oscillator algebra $\mathcal{A}_q(n)$ \cite{Macfarlane1989,Biedenharn1989,Floreanini1991} is defined as the unital associative algebra over $\CC$ generated by $n$ independent sets of $q$-oscillators $\{A_i^\pm$, $A_i^0\}$ verifying
\begin{equation}
\big[ A_i^0\,, A_i^\pm \big] = \pm A_i^\pm, \qquad \big[ A_i^- \,, A_i^+ \big] = q^{A_i^0}, \qquad A_i^- A_i^+ - q A_i^+ A_i^- = 1,\qquad i=1,\dots,n\,,
\end{equation}
and such that the commutators between elements with different indices $i$ are equal to zero.
The last two relations allow one to express $N_i = A_i^+ A_i^-$ in terms of $A_i^0$:
\begin{equation}
N_i = A_i^+ A_i^- = \frac{1-q^{A_i^0}}{1-q} =(A_i^0)_q.
\end{equation}
In the limit $q \to 1$, $A_i^0$ coincides with the usual number operator ${N}_i$. 

The $q$-oscillator algebra has the following representation on the space spanned by the standard occupancy number states $|n_1,\cdots,n_n\rangle = |n_1\rangle \otimes \cdots \otimes |n_n\rangle$ ($n_i \in \NN$):
\begin{equation}
A_i^0 |n_i\rangle = n_i |n_i\rangle , \qquad A_i^+ |n_i\rangle = \sqrt{\frac{1-q^{n_i+1}}{1-q}} |n_i+1\rangle , \qquad A_i^- |n_i\rangle = \sqrt{\frac{1-q^{n_i}}{1-q}} |n_i-1\rangle .
\label{eq:onbr}
\end{equation}
These commuting $q$-oscillators can now be used to build realizations of the algebras considered above. 

Firstly, the algebra $\mathfrak{o}_q(3)$ can be realized \`a la Schwinger in terms of two $q$-oscillators. More precisely, using the homomorphism $\chi : U_q(\mathfrak{sl}_2) \to \mathcal{A}_q(2)$ given by
\begin{equation}
\chi(j_0) = \half (A_1^0-A_2^0) , \qquad \chi(j_+) = q^{-\frac{1}{4}(A_1^0+A_2^0-1)} \, A_1^+A_2^- , \qquad \chi(j_-) = q^{-\frac{1}{4}(A_1^0+A_2^0-1)} \, A_1^-A_2^+ ,
\end{equation}
and the identification \eqref{eq:jiequit}, the following realization of $\mathfrak{o}_q(3)$ is obtained:
\begin{align}
& \chi(j_1) = \frac{iq^{\frac{1}{4}}}{q^{\frac{1}{2}}+q^{-\frac{1}{2}}} \, q^{-\frac{1}{2}A_2^0} \big( q^{\frac{1}{4}}A_1^-A_2^+ + q^{-\frac{1}{4}}A_1^+A_2^- \big) , \label{eq:j1schw} \\
& \chi(j_2) = \frac{q^{\frac{1}{4}}}{q^{\frac{1}{2}}+q^{-\frac{1}{2}}} \, q^{-\frac{1}{2}A_1^0} \big( q^{\frac{1}{4}}A_1^+A_2^- - q^{-\frac{1}{4}}A_1^-A_2^+ \big) . \label{eq:j2schw} 
\end{align}
Another key ingredient is the metaplectic realization of $U_q(\mathfrak{su}(1,1))$, which is given by the homomorphism $\mu : U_q(\mathfrak{su}(1,1)) \to \mathcal{A}_q(1)$:
\begin{align}\label{eq:metaplecticmap}
 \mu(J_0)=\mathscr{J}_0=\frac{1}{2}\left(A^{0}+\frac{1}{2}\right),\qquad \mu(J_{\pm})=\mathscr{J}_\pm=\frac{1}{[2]_{q^{1/2}}}(A^{\pm})^{2}.
\end{align}
One sees immediately that it is a $q$-deformation of the usual metaplectic representation of $\mathfrak{su}(1,1)$. 

Finally, we shall also use the realization of $\mathfrak{o}_{q^{1/2}}(4)$ in terms of $4$ $q$-oscillators which is provided by:
\begin{align}\label{eq:Lscriptiip1}
 \mathscr{L}_{i,i+1}= q^{-\frac{1}{2}A_i^0+\frac{1}{4}} \big( q^{\frac{1}{4}}A_i^+A_{i+1}^- - q^{-\frac{1}{4}}A_i^-A_{i+1}^+ \big),\qquad i=1,2,3.
\end{align}
One checks that the $\mathscr{L}_{i,i+1}$ indeed verify relations of the form \eqref{eq:soq4} but whose $q$'s have been replaced by $q^{1/2}$'s.
Furthermore, $\mathscr{L}_{12}$, $\mathscr{L}_{34}$ commute and hence generate a $\mathfrak{o}_{q^{1/2}}(2) \oplus \mathfrak{o}_{q^{1/2}}(2)$ subalgebra of $\mathfrak{o}_{q^{1/2}}(4)$.

\section{The commutant of $\mathfrak{o}_{q^{1/2}}(2) \oplus \mathfrak{o}_{q^{1/2}}(2)$ in the $q$-oscillator realization of $U_q(\mathfrak{u}(4))$ and the $q$-Higgs algebra}\label{sec:qhiggs}
It was shown in \cite{Frappat2018} that the Higgs algebra appears as the commutant of $\mathfrak{o}(2) \oplus \mathfrak{o}(2)$ in the universal enveloping algebra $U(\mathfrak{u}(4))$. This section aims to define the $q$-Higgs algebra through a $q$-analogue of this commutant picture.

We consider first the $\mathfrak{o}_{q^{1/2}}(2) \oplus \mathfrak{o}_{q^{1/2}}(2)$ subalgebra of $\mathfrak{o}_{q^{1/2}}(4)$ generated by $\mathscr{L}_{12}$ and $\mathscr{L}_{34}$, and look for its commutant in $U_q(\mathfrak{u}(4))$.


Introduce the following three operators
\begin{subequations}\label{eq:lmpmm}
\begin{align}
 M^+ &= \left( q^{A_2^0+\sfrac{1}{2}} (A_1^+)^2 + (A_2^+)^2 \right) \left( q^{A_4^0+\sfrac{1}{2}} (A_3^-)^2 + (A_4^-)^2 \right)\! , \label{eq:MP} \\
 M^- &= \left( q^{A_2^0+\sfrac{1}{2}} (A_1^-)^2 + (A_2^-)^2 \right) \left( q^{A_4^0+\sfrac{1}{2}} (A_3^+)^2 + (A_4^+)^2 \right)\! , \label{eq:MM} \\
 L\phantom{^{+!}} &= (A_1^0+A_2^0)-(A_3^0+A_4^0),
\end{align}
\end{subequations}
which commute with the generators $\mathscr{L}_{12}$ and $\mathscr{L}_{34}$ (in the limit $q \to 1$, $\mathscr{L}_{12}$ and $\mathscr{L}_{34}$ correspond to rotations in the $(1,2)$ and $(3,4)$ planes). 

One notes that each big parenthesis in the expression of the $M^{\pm}$ operators can actually be obtained by applying the coproduct of $U_q(\mathfrak{su}(1,1))$ to the $\mathscr{J}_\pm$ generators. 
Recalling that the bilinears of the form \mbox{$E_{ij}=A_{i}^{+}A_{j}^{-}$}, ~$i,j=1,2,3,4$ ~realize the $U_q(\mathfrak{u}(4))$ algebra \cite{Hayashi1990}, it can be observed that $M^{\pm}$, $L$ generate the non-trivial part of the commutant of $\mathfrak{o}_{q^{1/2}}(2) \oplus \mathfrak{o}_{q^{1/2}}(2)$ in the $q$-oscillator realization of $U_q(\mathfrak{u}(4))$.

It is immediate to see that $M^\pm$ and $L$ also commute with the central element
\begin{align}\label{eq:ham}
H = \sum_{i=1}^4 (A_i^0+\half).                                                                                                                                                                                                               \end{align}
One could ask how were the expressions for $L$, $M^{\pm}$ obtained. First, the operator $L$ obviously commutes with $\mathscr{L}_{12}$ and $\mathscr{L}_{34}$. Second, instead of obtaining the factors in $M^\pm$ from the coproduct one can look for elements $T^{\pm}$ in $\mathcal{A}_q(2)$ that commute with $\mathscr{L}_{12}$; this is most easily done ``on-shell'', that is, by solving \mbox{$[\mathscr{L}_{12},T^\pm] |n_1,n_2\rangle = 0$} ~for any two $q$-oscillator states. One thus arrives at
\begin{equation}\label{eq:Tpm}
T^\pm = q^{\alpha(A_1^0+A_2^0)} \big( q^{A_2^0+\frac{1}{2}} (A_1^\pm)^2 + (A_2^\pm)^2 \big),\qquad \alpha\in\mathbb{C}.
\end{equation}
Since $A_1^0+A_2^0 = \half(L+H)$, only the second factor of $T^\pm$ is relevant. The same is done with $\mathscr{L}_{34}$ on the direct product states $|n_3,n_4\rangle$. It is then clear that the only combinations of the operators \eqref{eq:Tpm} and their $(3,4)$ analogues that will belong to $U_q(\mathfrak{u}(4))$ are those occurring in $M^\pm$.


It now remains to determine the algebra formed by the three generators $M^\pm$ and $L$.
\begin{subequations}\label{eq:qhigtot}
\begin{prop}
The operators $M^\pm$ and $L$ have the following commutators:
\begin{align}\label{eq:qhig}
\begin{aligned}
\big[ L,M^\pm \big] &= \pm 4 M^\pm ,  \\
\big[ M^+,M^- \big] &= \frac{(1+q)}{q(1-q)^3} \; q^H \Big( (q+q^{-1})(q^L-q^{-L}) - 2 \big( q^{\frac{1}{2}H}+q^{-\frac{1}{2}H} \big)(q^{\frac{1}{2}L}-q^{-\frac{1}{2}L}) \Big) \\
& + \frac{(1+q)}{q^2(1-q)} \; q^H \Big( \big( q^{-\frac{1}{2}H} {\mathscr{L}_{12}}^2 + q^{\frac{1}{2}H} {\mathscr{L}_{34}}^2 \big) q^{\frac{1}{2}L} - \big( q^{\frac{1}{2}H} {\mathscr{L}_{12}}^2 + q^{-\frac{1}{2}H} {\mathscr{L}_{34}}^2 \big) q^{-\frac{1}{2}L} \Big) .
\end{aligned}
\end{align}
The elements ${\mathscr{L}_{12}}$, ${\mathscr{L}_{34}}$ and $H$, which are central, play the role of structure constants. We shall take these relations to define abstractly the (universal) $q$-Higgs algebra.
\end{prop}
\rmk
Alternatively, if one considers the generator $q^{\frac{1}{2}L}$ instead of $L$, the first set of relations in \eqref{eq:qhig} becomes
\begin{align}
 q^{\frac{1}{2}L}M^{\pm}=q^{\pm 2}M^{\pm}q^{\frac{1}{2}L}.\label{eq:qhig3}
\end{align}
\end{subequations}

\proof
The first relations of \eqref{eq:qhig} are obvious. The last relation is obtained by a direct computation in the $q$-oscillator algebra.
Starting with \eqref{eq:MP}--\eqref{eq:MM}, and using the identity $[a_1^+a_2^-,a_1^-a_2^+] = [a_1^+,a_1^-]a_2^+a_2^--a_1^+a_1^-[a_2^+,a_2^-]$ ~for $a_i^\pm = \Big( q^{A_{2i}^0+\sfrac{1}{2}} (A_{2i-1}^\pm)^2 + (A_{2i}^\pm)^2 \Big)$, one gets
\begin{align}
 \big[ M^+&,M^- \big] = \nonumber \\
& \Big[ q^{2A_{2}^0+1} \big[(A_{1}^+)^2 , (A_{1}^-)^2 \big] + \big[(A_{2}^+)^2 , (A_{2}^-)^2 \big] + q^{A_{2}^0+\sfrac{1}{2}} (1-q^2) \big( (A_{1}^+)^2 (A_{2}^-)^2 + q^{-2}(A_{1}^-)^2 (A_{2}^+)^2 \big)\Big] \nonumber \\ 
& \times \Big[ q^{2A_{4}^0+1} (A_{3}^+)^2 (A_{3}^-)^2 + (A_{4}^+)^2 (A_{4}^-)^2 + q^{A_{4}^0+\sfrac{1}{2}} \big( (A_{3}^+)^2 (A_{4}^-)^2 + q^{-2}(A_{3}^-)^2 (A_{4}^+)^2 \big) \Big] \nonumber \\ 
 - &\Big[ q^{2A_{4}^0+1} \big[(A_{3}^+)^2 , (A_{3}^-)^2 \big] + \big[(A_{4}^+)^2 , (A_{4}^-)^2 \big] + q^{A_{4}^0+\sfrac{1}{2}} (1-q^2) \big( (A_{3}^+)^2 (A_{4}^-)^2 + q^{-2}(A_{3}^-)^2 (A_{4}^+)^2 \big)\Big] \nonumber \\ 
& \times \Big[ q^{2A_{2}^0+1} (A_{1}^+)^2 (A_{1}^-)^2 + (A_{2}^+)^2 (A_{2}^-)^2 + q^{A_{2}^0+\sfrac{1}{2}} \big( (A_{1}^+)^2 (A_{2}^-)^2 + q^{-2}(A_{1}^-)^2 (A_{2}^+)^2 \big) \Big] .
\end{align}
Now, from the expression \eqref{eq:Lscriptiip1}, one obtains 
\begin{equation}
{\mathscr{L}_{12}}^2 = q^{-A_1^0+\sfrac{1}{2}} \Big( q(A_{1}^+)^2 (A_{2}^-)^2 + q^{-1}(A_{1}^-)^2 (A_{2}^+)^2 - q^{\sfrac{1}{2}} N_1  - q^{-\sfrac{1}{2}} N_2 - q(q^{\sfrac{1}{2}}+q^{-\sfrac{1}{2}}) N_1N_2 \Big)
\end{equation}
and a similar expression for ${\mathscr{L}_{34}}^2$ with the replacement $A_{1}^\bullet,A_{2}^\bullet \to A_{3}^\bullet,A_{4}^\bullet$.

Using the relations 
\begin{equation}
\big[ (A_{i}^+)^2,(A_{i}^-)^2 \big] = -(1+q) q^{A_i^0} \big((q+q^{-1})N_i+1\big)
\end{equation}
and
\begin{equation}
q (A_{i}^+)^2 (A_{i}^-)^2 = N_i^2-N_i ,
\end{equation}
and after some algebra, one is left with the following equation
\begin{align}
\begin{aligned}
 \big[ &M^+,M^- \big] =\\
&\Big[(1+q)q^{A^{0}_1+A^{0}_2-1}\left((1-q){\mathscr{L}_{12}}^2+\frac{q}{1-q}\left(q^{A^{0}_1+A^{0}_2}(1+q^{2})-2\right)\right)\Big]
\Big[q^{A^{0}_3+A^{0}_4-1}{\mathscr{L}_{34}}^2+[(A^{0}_3+A^{0}_4)_q]^{2}\Big]
\\
-&\Big[(1+q)q^{A^{0}_3+A^{0}_4-1}\left((1-q){\mathscr{L}_{34}}^2+\frac{q}{1-q}\left(q^{A^{0}_3+A^{0}_4}(1+q^{2})-2\right)\right)\Big] 
\Big[q^{A^{0}_1+A^{0}_2-1}{\mathscr{L}_{12}}^2+[(A^{0}_1+A^{0}_2)_q]^{2}\Big].
\end{aligned}
\end{align}
Expressing the $A_i^0$ generators in terms of $L$ and $H$, one finally obtains the desired commutation relation.

\begin{rmk}
In the limit $q \to 1$, noting that
\begin{align}
 \lim_{q\to1}\mathscr{L}_{12}=2i\,{\mathcal{L}}_{12},\qquad \lim_{q\to1}\mathscr{L}_{34}=2i\,{\mathcal{L}}_{34},\qquad\text{with}\quad \mathcal{L}_{jk}=-\frac{i}{2}\left(x_j\frac{\partial}{\partial x_k}-x_k\frac{\partial}{\partial x_j}\right),
\end{align}
one easily recovers from \eqref{eq:qhig} the commutation relations of the Higgs algebra \eqref{eq:higgsalg} in the form:
\begin{align}
\begin{aligned}\label{eq:higgs1}
\big[ L,M^\pm \big] &= \pm 4 M^\pm  , \\
\big[ M^+,M^- \big] &= -L^3 + \alpha_1 L + \alpha_2  ,
\end{aligned}
\end{align}
where $\alpha_1 = H^2+8({{\mathcal{L}}_{12}}^2+{{\mathcal{L}}_{34}}^2)-4$, and $\alpha_2 = -8H({{\mathcal{L}}_{12}}^2-{{\mathcal{L}}_{34}}^2)$.
\end{rmk}
Hence, the relations \eqref{eq:qhigtot} indeed define a $q$-deformation of the Higgs algebra.

\section{The Askey--Wilson algebra and an embedding into ${U_q(\mathfrak{su}(1,1))}^{\otimes2}$}\label{sec:aw}
We now indicate how (a special case of) the Askey--Wilson algebra can be embedded in the tensor product $U_q(\mathfrak{su}(1,1)) \otimes U_q(\mathfrak{su}(1,1))$. With $\Delta$ the coproduct of $U_q(\mathfrak{su}(1,1))$ given in \eqref{eq:coprodsu}, we can take
\begin{subequations}\label{eq:KiAW}
\begin{align}
K_1 &= \frac{1}{4}\;\frac{1-q^{J_0} \otimes q^{-J_0}}{1-q} , \label{eq:K1AW} \\
K_2 &= \frac{1}{2}\;\Delta(C) = \frac{1}{2} \bigg( C \otimes q^{2J_0} + q^{-2J_0} \otimes C + J_+ q^{-2J_0-1} \otimes J_- + J_- q^{-2J_0+1} \otimes J_+  \nonumber \\
&\qquad\qquad\quad + \frac{q^2+1}{(q^2-1)^2} \Big( q^{-2J_0} \otimes q^{2J_0} - 1 \otimes q^{2J_0} - q^{-2J_0} \otimes 1 + 1 \otimes 1 \Big) \bigg) , \label{eq:K2AW} 
\end{align}
where $C$ denotes the Casimir operator given in \eqref{eq:casimirsu}.

Defining $K_3 = \big[ K_1,K_2 \big]$, a direct calculation gives
\begin{equation}
K_3 = \frac{1}{8}\;(1+q^{-1}) \big( J_+ \otimes J_- - J_- \otimes J_+ \big) \big( q^{-J_0} \otimes q^{-J_0} \big). \label{eq:K3AW} 
\end{equation}
\end{subequations}
We now proceed to calculate the commutation relations of $K_1$, $K_2$, $K_3$. They are seen to take the form of the relations of the Askey-Wilson (AW) algebra which read
\begin{subequations}\label{eq:fullAW}
\begin{align}
\big[ K_1, K_2 \big] &= K_3 , \label{eq:AW1} \\
\big[ K_2, K_3 \big] &= r K_2 K_1 K_2 + \xi_1 \{ K_1,K_2 \} + \xi_2 K_2^2 + \xi_3 K_2 + \xi_4 K_1 + \xi_5 , \label{eq:AW2} \\
\big[ K_3, K_1 \big] &= r K_1 K_2 K_1 + \xi_1 K_1^2 + \xi_2 \{ K_1,K_2 \} + \xi_3 K_1 + \xi_6 K_2 + \xi_7 , \label{eq:AW3} 
\end{align}
\end{subequations}
where $r$ is as in \eqref{eq:params} below and $\xi_1,\dots, \xi_7$ are arbitrary in the generic AW situation.
 
After a rather cumbersome calculation, using the expressions \eqref{eq:KiAW} for the $K_i$'s as well as the commutation relations \eqref{eq:relcomsu}, one finds that the $K_i$'s indeed obey the relations \eqref{eq:fullAW} with the following specific expressions for the parameters: 
\begin{align}\label{eq:params}
\begin{aligned}
& r = -(q-q^{-1})^2 , \qquad \xi_1 = \frac{1+q^{-2}}{2} , \qquad \xi_2 = \frac{(1+q)^{2}(1-q)}{4q^2} , \qquad \xi_3 = 4(q-1)\,\xi_7,\\
& \xi_4 = 0 , \qquad \xi_5 = -\frac{(1+q)(1+q^2)}{16q^3} (C^{(1)}-C^{(2)}) \big[ J_0^{(12)} \big]_{q} , \qquad \xi_6 = -\frac{(1+q)^2}{16q^2} , \\
& \xi_7 = \frac{(1+q)^2}{32q^2} \left( C^{(1)} q^{J_0^{(12)}} + C^{(2)} q^{-J_0^{(12)}} - (1+q^{-2}) \big[ \half J_0^{(12)} \big]_{q}^2 \right),  
\end{aligned}
\end{align}
where $C^{(1)} = C \otimes 1$ and $C^{(2)} = 1 \otimes C$ are respectively the Casimir operators in the spaces $1$ and $2$ of the tensor product, and $J_0^{(12)} = \Delta(J_0)$. These quantities $C^{(i)}$ and $J_0^{(12)}$ commute with $K_1$, $K_2$ and $K_3$ and we hence have a version of \eqref{eq:fullAW} that is centrally extended.

Since there are only three independant quantities entering the $\xi_i$'s (there are four in the general case), 
we conclude that the $K_1$, $K_2$, $K_3$ generate a specialization of the Askey--Wilson algebra. One checks that in the limit $q\to1$, the parameters $r$, $\xi_2$, $\xi_3$ vanish and one recovers the Hahn algebra.
The standard $q$-Hahn algebra is obtained from the Askey-Wilson algebra by setting for instance $\xi_1=0$ in \eqref{eq:fullAW}. The algebra satisfied by $K_1$, $K_2$ and $K_3$ is actually isomorphic to the $q$-Hahn algebra as the standard form of the latter \cite{Zhedanov1991,Baseilhac2018} is obtained by taking $K_2 = \widetilde{K}_2 - \xi_1/r$. The limit $q\to1$ is singular however if we adopt this presentation.

\section{The $q$-Higgs algebra, the Askey--Wilson algebra, and the dual pair $\big( \mathfrak{o}_{q^{1/2}}(4)\,,U_q(\mathfrak{su}(1,1)) \big)$}\label{sec:dualpair}
We shall explain in this section how the $q$-Higgs algebra obtained as a commutant and the special Askey-Wilson algebra found from the embedding just described are connected through Howe duality and are in fact isomorphic.

Take $4$ metaplectic representations defined as in \eqref{eq:metaplecticmap}. We will add them first pairwise using the $U_q(\mathfrak{su}(1,1))$ coproduct \eqref{eq:coprodsu}:
\begin{align}\label{eq:embreal}
\mathscr{J}_0^{(2i-1,2i)} = \half \big( A_{2i-1}^0 + A_{2i}^0 + 1 \big), \qquad \mathscr{J}_\pm^{(2i-1,2i)} = \frac{1}{[2]}_{q^{1/2}} \Big( q^{A_{2i}^0+\sfrac{1}{2}} (A_{2i-1}^+)^2 + (A_{2i}^+)^2 \Big),\qquad i=1,2.
\end{align}
and then using the coproduct once more will form
\begin{align}
\mathscr{J}^{(1234)}_0 = \mathscr{J}^{(12)}_0 + \mathscr{J}^{(34)}_0 \qquad \text{and} \qquad
\mathscr{J}^{(1234)}_\pm = \mathscr{J}^{(12)}_\pm q^{2\!\mathscr{J}^{(34)}_0} + \mathscr{J}^{(34)}_\pm.
\end{align}
Mindful of Section \ref{sec:qhiggs}, it is immediate to check that
\begin{align}\label{eq:oq4su11comm}
 [\mathscr{L}_{i,i+1},\mathscr{J}_{\bullet}^{(1234)}]=0,\qquad i=1,2,3,
\end{align}
where the $\mathscr{L}_{i,i+1}$ are defined as in \eqref{eq:Lscriptiip1}. Let us stress that \eqref{eq:oq4su11comm} makes the key statement that the algebras $U_q(\mathfrak{su}(1,1))$ and $\mathfrak{o}_{q^{1/2}}(4)$ are mutually commuting in the $q$-oscillator realization.

It has been shown \cite{Noumi1996} that $\mathfrak{o}_{q^{1/2}}(4)$ and $U_q(\mathfrak{su}(1,1))$ actually form a Howe dual pair. (They constitute precisely the quantum analogue of the classical pair $\left(\mathfrak{o}(4),\mathfrak{su}(1,1)\right)$ which was used in the analysis of the Higgs and Hahn algebras \cite{Frappat2018}.) This means that their representations can be connected through their Casimirs. We will now proceed to indicate explicitly how this is realized. 

To that end, we first put the $\mathscr{J}_{\bullet}^{(2i-1,2i)}$ in correspondance with the $J_{\bullet}$ from Section \ref{sec:aw}. Let us focus on the coproduct embeddings \eqref{eq:embreal}. As each pairing of $U_q(\mathfrak{su}(1,1))$ in the spaces $(1,2)$ and $(3,4)$ gives a copy of $U_q(\mathfrak{su}(1,1))$, we can embed the specialization of the Askey--Wilson algebra of Section \ref{sec:aw} into these two copies of $U_q(\mathfrak{su}(1,1))$.

Indeed, upon substitution of \eqref{eq:embreal} into equations \eqref{eq:KiAW} for $K_1$, $K_2$, $K_3$, we obtain the following $q$-oscillator realization of the special Askey--Wilson algebra:
\begin{subequations}
\begin{align}
\mathscr{K}_1 &= \frac{1}{4} \; \frac{1-q^{\frac{1}{2}L}}{1-q} , \\
\mathscr{K}_2 &= \frac{1}{2} \Delta^{(3)}(C) = \frac{1}{2} \left( \big( \mathscr{C}^{(1)} q^{\half H} + \mathscr{C}^{(2)} q^{-\half H} \big) q^{-\half L} + (1+q^{-2}) \, q^{-\half L} \big[ \sfrac{L+H}{4} \big]_q \big[ \sfrac{L-H}{4} \big]_q \right. \nonumber \\
& \hspace{5.7em}+ \left. \frac{q}{(1+q)^2} \big( q^{-1} M_+ + qM_-\big) q^{-\half(H+L)}\right), \\
\mathscr{K}_3 &= \frac{1}{8(1+q)} \; \big( M_+ - M_- \big) q^{-\half H} .
\end{align}
\end{subequations}
where $\Delta^{(n)}(x)=(\id^{\otimes (n-1)}\otimes\Delta)\Delta^{(n-1)}(x)$, $\Delta^{(1)}=\Delta\,$, $\Delta^{(0)}=\id$; the generators $M_\pm$, $L$, $H$ correspond to those given in \eqref{eq:lmpmm} and \eqref{eq:ham} respectively and can alternatively be expressed as
\begin{align}
M_\pm &= ([2]_{q^{1/2}})^2 \mathscr{J}^{(12)}_\pm \mathscr{J}^{(34)}_\mp,\\
\half L &= \mathscr{J}^{(12)}_0 - \mathscr{J}^{(34)}_0,\\
\half H &= \mathscr{J}^{(12)}_0 + \mathscr{J}^{(34)}_0=\Delta^{(3)}(\mathscr{J}_0).
\end{align}
By construction the $\mathscr{K}_1$, $\mathscr{K}_2$, $\mathscr{K}_3$ obey the relations of the special Askey--Wilson algebra \eqref{eq:fullAW} with the parameters \eqref{eq:params}.

Also note that the quantities $\mathscr{C}^{(1)}$ and $\mathscr{C}^{(2)}$ are the images of the Casimir operators $C^{(1)}$ and $C^{(2)}$ and that they are directly related to the ${\mathscr{L}_{12}}$ and ${\mathscr{L}_{34}}$ by
\begin{align}
\mathscr{C}^{(1)} = \frac{1}{(1+q)^{2}}\left({\mathscr{L}_{12}}^2+1\right), \qquad  \mathscr{C}^{(2)} = \frac{1}{(1+q)^{2}}\left({\mathscr{L}_{34}}^2+1\right).
\end{align}
In view of this, it is now evident that in the $q$-oscillator realization, the generators of the special Askey--Wilson algebra are expressible in terms of those of the $q$-Higgs algebra, and vice-versa. Hence these two algebras are isomorphic, as in the $q\to1$ case.

To wrap things up, let us point out that the two Casimirs of $\mathfrak{o}_{q^{1/2}}(4)$ given in \eqref{eq:casimirsoq4} have a direct interpretation in this $q$-oscillator framework.

The first Casimir of $\mathfrak{o}_{q^{1/2}}(4)$, denoted $C_4$, corresponds to the total Casimir of the quadruple tensor product of $U_q(\mathfrak{su}(1,1))$:
\begin{align}
 C_4=\left(
q^{-1} {\mathscr{L}_{12}}^2 + {\mathscr{L}_{23}}^2 + q {\mathscr{L}_{34}}^2 + q^{-\frac{1}{2}} \mathscr{L}_{13}^+\,\mathscr{L}_{13}^- + q^{\frac{1}{2}} \mathscr{L}_{24}^+\,\mathscr{L}_{24}^- + \mathscr{L}_{14}^+\,\mathscr{L}_{14}^- \right)=\frac{(1+q)^{2}}{2} \Delta^{(3)}(C).
\end{align}
This is precisely the pairing of the Casimirs of $\mathfrak{o}_{q^{1/2}}(4)$ and $U_q(\mathfrak{su}(1,1))$ that follows from the Howe duality.

The second Casimir of $\mathfrak{o}_{q^{1/2}}(4)$, denoted $C'_4$, is identically zero in the $q$-oscillator realization:
\begin{align}
 C'_4=q^{-\frac{1}{2}}\mathscr{L}_{12}\mathscr{L}_{34}-\mathscr{L}_{13}^{+}\mathscr{L}_{24}^{+}+q^{\frac{1}{2}}\mathscr{L}_{23}\mathscr{L}_{14}^{+}=0.
\end{align}
It can be seen as the $q$-analogue of the usual relation between the angular momenta, see for instance $(4.1)$ in \cite{Feigin2015}: $M_{12}M_{34}+M_{13}M_{42}+M_{14}M_{23}=0$.

Let us mention in closing this section that the $q\to1$ limit of the above yields straightforwardly the duality presented in \cite{Frappat2018} between the Higgs or the Hahn algebras viewed as a commutant in $U(\mathfrak{u}(4))$ or embedded in $U(\mathfrak{su}(1,1))\otimes U(\mathfrak{su}(1,1))$.

\section{Conclusion}
Summing up, we have introduced a $q$-analogue of the Higgs algebra by looking for the commutant of a \mbox{$\mathfrak{o}_{q^{1/2}}(2)\oplus\mathfrak{o}_{q^{1/2}}(2)$} subalgebra of $\mathfrak{o}_{q^{1/2}}(4)$ in the $q$-oscillator representation of $U_q(\mathfrak{u}(4))$. This algebra was then seen to be isomorphic to a special case of the Askey--Wilson algebra (itself isomorphic to the standard $q$-deformation of the Hahn algebra) which has an embedding in $U_q(\mathfrak{su}(1,1))\otimes U_q(\mathfrak{su}(1,1))$. The Howe dual pair $\big(\mathfrak{o}_{q^{1/2}}(4),U_q(\mathfrak{su}(1,1))\big)$ was then invoked as the reason behind this double picture.

The $q$-oscillator realization in which $\mathfrak{o}_{q^{1/2}}(4)$ and $U_q(\mathfrak{su}(1,1))$ commute can be generalized easily for $\mathfrak{o}_{q^{1/2}}(n)$ with $n$ arbitrary. It is known that $\big(\mathfrak{o}_{q^{1/2}}(n),U_q(\mathfrak{su}(1,1))\big)$ is a dual pair \cite{Noumi1996}. This opens up the door to the study of the full Askey--Wilson algebra. We hypothesize that it should be possible to obtain this algebra as the commutant of a $\mathfrak{o}_{q^{1/2}}(2)\oplus\mathfrak{o}_{q^{1/2}}(2)\oplus\mathfrak{o}_{q^{1/2}}(2)$ subalgebra of $\mathfrak{o}_{q^{1/2}}(6)$ in $U_q(\mathfrak{u}(6))$ in this $q$-oscillator representation. It would be also of high interest to see if the higher rank Askey--Wilson algebras \cite{Post2017,DeBie2018} could be obtained in a similar fashion.

It should moreover be mentioned that the dual pair $\big(\mathfrak{o}_{q^{1/2}}(n),U_q(\mathfrak{su}(1,1))\big)$ was analyzed in \cite{Noumi1996} in a $q$-commuting variable framework. It would be quite interesting to see if some sort of dimensional reduction in $q$-commuting variables could be performed to obtain a $q$-analogue of the superintegrable model on the $n$-sphere \cite{Kalnins2007}. We hope to address all these questions in the near future.

\subsection*{Acknowledgments}
LV wishes to acknowledge the hospitality of the CNRS and of the LAPTh in Annecy where part of this work was done. 
ER and LF are also thankful to the Centre de Recherches Mathématiques (CRM) for supporting their visits to Montreal in the course of this investigation. 
JG holds an Alexander-Graham-Bell scholarship from the Natural Science and Engineering Research Council (NSERC) of Canada. 
The research of LV is supported in part by a Discovery Grant from NSERC.

\bibliographystyle{unsrtinurl} 
\bibliography{citationsqHiggs.bib}

\end{document}